\documentclass[prl,%
twocolumn,%
tightenlines,shownopacs,%
nofootinbib,%
superscriptaddress,%
amsfonts,amsmath]{revtex4}

\usepackage{graphicx}
\usepackage{amsmath}

\begin{document}

\title{Instability of black holes in massive gravity}

\author{Eugeny~Babichev} \email{eugeny.babichev@th.u-psud.fr}
\affiliation{Laboratoire de Physique Th\'eorique d'Orsay,
B\^atiment 210, Universit\'e Paris-Sud 11,
F-91405 Orsay Cedex, France}

\author{Alessandro Fabbri} \email{afabbri@ific.uv.es}
\affiliation{Museo Storico della Fisica e Centro Studi e Ricerche ’Enrico Fermi’, Piazza del Viminale 1, 00184 Roma, Italy; \\ \noindent Dipartimento di Fisica dell’Universit\`a di Bologna,
Via Irnerio 46, 40126 Bologna, Italy; \\ \noindent Departamento de F\'isica Te\'orica and IFIC, Universidad de Valencia-CSIC, C. Dr. Moliner 50, 46100 Burjassot, Spain}

\begin{abstract}
We show that linear perturbations around the simplest black hole solution of massive bi-gravity theories, the bi-Schwarzschild solution, exhibit an unstable mode featuring the Gregory-Laflamme instability of higher-dimensional black strings.  
The result is obtained for the massive gravity theory which is free from the Boulware-Deser ghost, as well as for its extension with two dynamical metrics.
These results may indicate that static black holes in massive gravity do not exist.
For graviton mass of order of Hubble scale, however, the instability timescale is of order of the Hubble time.  
\end{abstract}

\date{\today}

%\pacs{04.50.Kd [Modified theories of gravity], 14.70.Kv [Gravitons], 04.50.Gh [Higher-dimensional black holes, black strings, and related objects]}

\maketitle

Massive gravity is an alternative to General Relativity (GR) in which the graviton is supposed to gain mass. 
There are several motivations behind the study of massive gravity:
theoretical curiosity to construct a gravitation theory with a massive graviton, 
to add a new model in the zoo of alternative theories in order to test better GR compared to alternatives, 
explain present day cosmological acceleration of the universe in a more ``natural'' way than introduction of the cosmological constant,
resolve the vacuum energy problem by hoping that the graviton becomes weaker {\it \`a la} Yukawa, etc. 

Massive gravity has started from the works by Fierz and Pauli \cite{Fierz:1939ix}, 
who studied Lorentz invariant massive spin 2 theory (hereafter referred to as FP), described by a quadratic action.
In contrast to GR this model has five propagating degrees of freedom and 
it is excluded by Solar system experiments due to an extra force associated with the helicity-0 mode.
This extra force does not vanish in the limit of zero graviton mass, which is a consequence of the van Dam-Veltman-Zakharov discontinuity~\cite{vanDam:1970vg}.
It was, however, conjectured by Vainshtein in~\cite{Vainshtein:1972sx} that GR can be restored at nonlinear level 
for nonlinear massive gravity theories. 
A nonlinear completion of FP can be achieved by introducing a second metric~\cite{Isham:1971gm},
which is needed to construct a mass term (in the original linear FP model 
the role of the second metric is played by the flat metric 
coinciding with the background solution for the first metric). 
Whether the Vainshtein mechanism indeed works, 
namely whether there are solutions around matter sources approximating GR inside some radius (the Vainshtein radius)
that continuously join an asymptotically flat solution of the linearized FP,
was an unresolved question for a long time.
Only recently it was shown that for a large class of nonlinear FP models the Vainshtein mechanism indeed works \cite{Babichev:2009us}. 

Another problem of massive gravity is the appearance of a ghost propagating degree of freedom at nonlinear level 
in generic nonlinear FP theory \cite{Boulware:1973my}, the so called Boulware-Deser ghost.
This problem has been also resolved only very recently  by de Rham, Gabadadze and Tolley~\cite{deRham:2010ik}  (dRGT theory), who found 
a two-parameter family of nonlinear FP theories 
free from the Boulware-Deser ghost.
Similar to the general nonlinear FP model, the Vainshtein mechanism also operates in this model~\cite{Koyama:2011xz,Volkov:2012wp}.
The dRGT model can be generalized by promoting the second metric to be dynamical~\cite{Hassan:2011zd}.
Although the dRGT theory and its extension are disputed from different angles~\cite{Chamseddine:2013lid,Kluson:2013cy,Deser:2012qx,Izumi:2013poa,ArkaniHamed:2002sp}, these are the only candidate to be pathology-free nonlinear FP theory so far.
Therefore, it is certainly worthwhile to study their different applications and physical consequences.

In this paper we show that the simplest black hole solutions 
in both dRGT theory and its extension are unstable.  Namely, the bi-gravity solutions with two Schwarzschild metrics simultaneously diagonal exhibit an unstable mode.
The result is valid for bi-Schwarzschild solutions with both equal and proportional metrics.

One can divide black hole solutions in bi-gravity theories into two types. 
The first one corresponds to solutions in which the metrics are not simultaneously bi-diagonal, 
i.e. if one in some coordinates is diagonal, the other metric is not. 
It is possible to find some classes of solutions analytically, \cite{Koyama:2011yg,Comelli:2011wq,Volkov:2012wp} [see also recent reviews~\cite{Volkov:2013roa} and references within].
Such solutions typically have a very non-trivial global structure, similar to those in the general nonlinear FP theory~\cite{Blas:2005yk} and
it is not clear if they are of physical relevance, in particular 
it may be difficult to form this type of solutions during gravitational collapse 
(if the result of gravitational collapse in massive gravity is a black hole).

On the other hand, spherically symmetric static bi-diagonal solutions, 
which seem to be more natural outcome of gravitational collapse 
(at least from configurations with smooth sources featuring the Vainshtein behavior),
must share the same Killing horizon~\cite{Deffayet:2011rh}, if one assumes non-singular solutions.
Therefore the variety of bi-diagonal solutions is highly restricted if the non-singularity of both metrics is assumed.

It is, however, not difficult to extend the known solutions in GR to bi-gravity, by simply identifying 
the two metics with a solution for  GR. 
This ensures that the interaction mass term is zero and the equations of motion are satisfied for both the dRGT model and its extension.
For example, the analog of the simplest black hole solution in GR, the Schwarzschild solution, is the bi-Schwarzschild solution in bi-gravity.
In this paper we will show explicitly that its linear perturbations satisfy the same equations studied by Gregory and Laflamme (GL) for the relevant four dimensional perturbations of higher-dimensional black string solutions~\cite{Gregory:1993vy}, 
featuring the well known GL instability. 

The action for the extension of the dRGT model with two dynamical metrics reads~\cite{deRham:2010ik,Hassan:2011zd},
\begin{equation}\label{ACTION}
	S = M^2_P\int d^4 x\sqrt{-g}\left( \frac{R_g}{2} - m^2 \mathcal{U} \right)
	 + \frac{\kappa M^2_P}{2}\int d^4 x \sqrt{-f} \mathcal{R}_f,
\end{equation}
where $M_P$ is the Planck mass, $R_g$ and $\mathcal{R}_f$ are the Einstein-Hilbert terms for the $g$ and $f$ metrics correspondingly,
$\kappa$ is a dimensionless parameter and  
$\mathcal{U}$ is the interaction mass term, 
\begin{equation}\label{U}
	\mathcal{U} = \mathcal{U}_2 + \alpha_3 \mathcal{U}_3 + \alpha_4 \mathcal{U}_4.
\end{equation}
The interaction can be written in terms of the matrix $\mathcal{K}^\mu_\nu$, 
$\mathcal{K}^\mu_\nu \equiv \delta^\mu_\nu - \sqrt{g^{\mu\alpha}f_{\alpha\nu}}$, where the square root of a matrix $\mathcal{A}$ 
is understood as $(\sqrt{\mathcal{A}})^\mu_\alpha (\sqrt{\mathcal{A}})^\alpha_\nu = \mathcal{A}^\mu_\nu$.
The pieces of the ghost-free interaction term (\ref{U}) can be cast in the following form,
\begin{equation}\label{U234}	
	\begin{aligned}
	\mathcal{U}_2 &= -\frac{1}{2!}\left( \left(\mathcal{K^\mu_\mu}\right)^2 - \mathcal{K}^\mu_\nu\mathcal{K}^\nu_\mu \right), \\
	\mathcal{U}_3 &= \frac{1}{3!}\epsilon_{\mu\nu\rho\sigma}\epsilon^{\alpha\beta\gamma\sigma}
		\mathcal{K}^\mu_\alpha \mathcal{K}^\nu_\beta \mathcal{K}^\rho_\gamma,\\
		\mathcal{U}_4 &= \frac{1}{4!}\epsilon_{\mu\nu\rho\sigma}\epsilon^{\alpha\beta\gamma\delta}
		\mathcal{K}^\mu_\alpha \mathcal{K}^\nu_\beta \mathcal{K}^\rho_\gamma \mathcal{K}^\sigma_\delta.
	\end{aligned}
\end{equation}
Varying the action with respect to $g_{\mu\nu}$ and $f_{\mu\nu}$ gives correspondingly, in the vacuum,
\begin{equation}
	G_{\mu\nu} = m^2 T^\mathcal{U}_{\mu\nu},\; 	
	\mathcal{G}_{\mu\nu} =  \frac{\sqrt{-g}}{\sqrt{-f}}\frac{m^2}{\kappa} \mathcal{T}^{\mathcal{U}}_{\mu\nu},
	\label{EOMS}
\end{equation}
where $G_{\mu\nu}$ and $\mathcal{G}_{\mu\nu}$ are the Einstein tensors for $g_{\mu\nu}$ and  $f_{\mu\nu}$ correspondingly, 
and $T^\mathcal{U}_{\mu\nu}$ and $\mathcal{T}^{\mathcal{U}}_{\mu\nu}$ come from the interaction term, 
\begin{equation}\label{T}
\begin{aligned}
	T^\mathcal{U}_{\mu\nu} &= 
	-g_{\mu\beta}\left(\delta^\beta_\alpha-\mathcal{K}^\beta_\alpha\right) \left(\mathcal{K}^\alpha_\nu -\mathcal{K}^\lambda_\lambda\delta^\alpha_\nu\right) + \mathcal{O}(\mathcal{K}^3),\\
	\mathcal{T}^\mathcal{U}_{\mu\nu} &= - T^\mathcal{U}_{\mu\nu} + \mathcal{O}(\mathcal{K}^3),
\end{aligned}
\end{equation}
where the higher order terms in $\mathcal{K}$ we collectively denoted as $\mathcal{O}(\mathcal{K}^2)$. 

Note that the dRGT model for one dynamical metric is described by~(\ref{ACTION}) without the last term, 
defining dynamics of the metric $f_{\mu\nu}$.
In this case the metric $f_{\mu\nu}$ becomes non-dynamical and  one should solve only the first equations in (\ref{EOMS}).

If both metrics are identical, $g_{\mu\nu} = f_{\mu\nu}$, then $\mathcal{K} = 0$, and hence 
$\mathcal{T}^\mathcal{U}_{\mu\nu}=T^\mathcal{U}_{\mu\nu} =0$.
Therefore any vacuum GR solution $g_{\mu\nu}$ is also a solution for the bi-gravity theory provided that  $f_{\mu\nu}=g_{\mu\nu}$.
In particular, the simplest black hole solution in bi-gravity theory is obtained by identifying 
the line element for each of the metrics with the Schwarzschild one,
\begin{equation}\label{S}
	ds^2 = -\left(1-\frac{r_S}{r}\right)dt^2 + \left(1-\frac{r_S}{r}\right)^{-1}dr^2 + r^2 d\Omega^2,
\end{equation}
where $r_S$ is the Schwarzschild radius. 
Now let us consider perturbations on top of the bi-Schwarzschild solution, so that we write 
\begin{equation}\label{METRICPERT}
g_{\mu\nu} = g^{(0)}_{\mu\nu} + h_{\mu\nu}, \quad f_{\mu\nu} = f^{(0)}_{\mu\nu} + \tilde{h}_{\mu\nu},
\end{equation}
where $g^{(0)}_{\mu\nu} = f^{(0)}_{\mu\nu}$ is given by (\ref{S}).
It is not difficult to see that $g^{\mu\alpha}f_{\alpha\nu} = \delta^\mu_\nu + \left(\tilde{h}^{\mu}_{\nu} -h^{\mu}_{\nu}  \right)$ and
\begin{equation}\label{K}
\mathcal{K}^{\mu}_\nu = \frac12 h^{(-)\mu}_{\phantom{(-)}\nu} + \mathcal{O}(h^2),
\end{equation}
where we introduced the notations $h^{(-)\mu}_{\phantom{(\pm)}\nu}= h^{\mu}_{\nu} - \tilde{h}^{\mu}_{\nu} $ and 
$h^{(+)\mu}_{\phantom{(\pm)}\nu}= h^{\mu}_{\nu} + \kappa\tilde{h}^{\mu}_{\nu} $.
Linearization of (\ref{EOMS}) yields,
\begin{eqnarray}
\mathcal{E}^{\alpha\beta}_{\mu\nu}h_{\alpha\beta} +\frac{ m^2}{2}\left(h^{(-)}_{\mu\nu} -  g_{\mu\nu}h^{(-)} \right) &=0,  \label{PERTg}\\
\mathcal{E}^{\alpha\beta}_{\mu\nu}\tilde h_{\alpha\beta} -\frac{ m^2}{2\kappa}\left(h^{(-)}_{\mu\nu} -  g_{\mu\nu}h^{(-)} \right) &= 0 \label{PERTf}
\end{eqnarray}
where
\begin{equation}
\begin{aligned}
\mathcal{E}^{\alpha\beta}_{\mu\nu}h_{\alpha\beta}  &= 
-\frac12\left( \nabla_{\mu}\nabla_{\nu}h - \nabla_\nu \nabla_\sigma h^\sigma_\mu-  \nabla_\mu \nabla_\sigma h^\sigma_\nu \right.\\
 &+  \left.  \Box h_{\mu\nu}  - g_{\mu\nu} \Box h + g_{\mu\nu}\nabla_{\alpha}\nabla_{\beta}h^{\alpha\beta} + 2 R^{\sigma\phantom{\mu}\lambda}_{\phantom{\mu}\mu\phantom{\mu}\nu} h_{\lambda_\sigma}  \right)\nonumber
\end{aligned}
\end{equation}
is the linearized Einstein tensor on top of the metric $g_{\mu\nu}$ in the vacuum,  $R_{\mu\nu} = R =0 $.
Combining equations (\ref{PERTg}) and (\ref{PERTf}) we obtain, 
\begin{eqnarray}
\mathcal{E}^{\alpha\beta}_{\mu\nu} h^{(-)}_{\alpha\beta}  &+& \frac{m'^{2}}{2}\left(h^{(-)}_{\mu\nu} -  g_{\mu\nu}h^{(-)} \right) = 0,  \label{PERTMASSa}\\
\mathcal{E}^{\alpha\beta}_{\mu\nu} h^{(+)}_{\alpha\beta}  &=& 0,\label{PERTMASSb}
\end{eqnarray}
where $m'\equiv m\sqrt{1+1/\kappa}$.
Therefore $h^{(-)}_{\mu\nu}$ is a massive and $h^{(+)}_{\mu\nu}$ is a massless spin-2 perturbation.
Acting on~(\ref{PERTMASSa}) with $\nabla^\nu$ gives $\nabla^\nu h^{(-)}_{\mu\nu}=\nabla_\mu h^{(-)}$ 
and hence $\nabla^\nu\nabla^\mu h^{(-)}_{\mu\nu}=\Box h^{(-)}$. Using this last relation in the trace of~(\ref{PERTMASSa})
one finds the conditions to be satisfied by $h^{(-)}$:
\begin{equation}\label{COND}
\nabla^\nu h^{(-)}_{\mu\nu}  =  h^{(-)} = 0.
\end{equation}
The equation for the massive spin-2 perturbation (\ref{PERTMASSa})  with the conditions  (\ref{COND}) can be rewritten as
\begin{equation}\label{GLEQ}
\Box h^{(-)}_{\mu\nu} + 2 R^{\sigma\phantom{\mu}\lambda}_{\phantom{\mu}\mu\phantom{\mu}\nu} h^{(-)}_{\lambda_\sigma} = m'^2 h^{(-)}_{\mu\nu}.
\end{equation}
Note that under the infinitesimal transformation associated with the change of coordinates $x^\mu \to x^\mu + \xi^\mu$
the massive spin-2 perturbation $h^{(-)}_{\mu\nu}$ does not change, 
while for the massless spin-2 perturbation $h^{(+)}_{\mu\nu}$ this is the gauge transformation (which leaves the equations of motion unchanged)
as for linearized GR. 

So far we considered the bi-dynamical extension of the dRGT model, where the two metrics are dynamical. 
In the context of the original dRGT model (with one dynamical metric, the other fixed and equal to Schwarzschild), the same type of equation for the massive perturbations is valid.
Indeed, in this case only perturbations of the dynamical metric are present,  $h_{\mu\nu}$, and they obey 
Eq.~(\ref{PERTMASSa}) with $ h^{(-)}_{\mu\nu} \to  h_{\mu\nu} $  and $\kappa\to \infty$.
Note that in the context of the original dRGT model any choice of the fixed fiducial metric corresponds to a different theory, 
and therefore the particular bi-Schwarzschild solutions we discuss here are very peculiar. 
However, these are the only regular bi-diagonal solutions~\cite{Deffayet:2011rh}.
We will comment about non-bi-diagonal solutions later on.

The solutions to Equation~(\ref{GLEQ}) with the conditions  (\ref{COND}) have already been studied in the context of the GL instability  ~\cite{Gregory:1993vy}. In its simplest setting,  linear perturbations around  the five-dimensional black string
\begin{equation}
ds^2= -\left(1-\frac{r_S}{r}\right)dt^2 + \left(1-\frac{r_S}{r}\right)^{-1}dr^2 + r^2 d\Omega^2 +dz^2
\end{equation}
were considered. Performing a Fourier decomposition around the infinite 5th flat dimension,   
it was shown that the relevant perturbations are the 4D ones 
$\int dk 
e^{ikz}h_{\mu\nu}^{(4)}$ 
(no instability was found in the $h_{\mu z}$ and $h_{zz}$ components), 
and that $h_{\mu\nu}^{(4)}$ satisfy the same massive spin two Equations~(\ref{GLEQ}) with $m'^2=k^2$ in the Schwarzschild background. Unstable ($\Omega>0$) spherically symmetric modes 
\begin{equation}\label{pert4D}
h_{\mu\nu}^{(4)}=e^{\Omega t}\left(
  \begin{array}{cccc}
   H_{tt}(r) & H_{tr}(r)  & 0 & 0 \\
   H_{tr}(r) & H_{rr}(r)  & 0 & 0 \\
    0 & 0 & K(r) & 0 \\
   0 & 0 & 0 & K(r)\sin^2\theta \\
  \end{array}\right)\ 
\end{equation}
regular at the future event horizon were found within the range \begin{equation}\label{inst} 0<m'<\frac{\mathcal{O}(1)}{r_S}\ .\end{equation} 
Based on these results we conclude that the bi-Schwarzschild massive gravity solutions are unstable too provided $m' = m\sqrt{1+1/\kappa}$ satisfies (\ref{inst}).

It is possible to extend these results to the case of proportional metrics, $f_{\mu\nu} =\omega^2 g_{\mu\nu}$. 
By appropriate choice of  $\alpha_3$, $\alpha_4$ and $\omega$, one can obtain the bi-Schwarzschild solution for both the dRGT theory and its extension.
The requirement of asymptotic flatness yields $\omega$ as a function of $\alpha_3$ and $\alpha_4$ for the dRGT model, 
while for its extension the asymptotic bi-flatness leaves only one-parameter family of solutions for $\omega$.
One can check that the mass term in the perturbation equation(s) 
is merely modified by a factor  depending on $\alpha_3$, $\alpha_4$ and $\omega$\footnote{Note that in this case not only the linear terms in $\mathcal{K}$ in Eq.~(\ref{T}) but also the rest contribute to the mass term for the perturbations.}.
The equations for perturbations remain  the same, up to the redefinition of the mass and therefore the 
instability is present in those solutions too if the inequality (\ref{inst}) is satisfied\footnote{There is a specific choice of parameters which gives zero mass term for the perturbation(s). 
This corresponds to a solution with infinitely strongly coupled perturbations. We do not consider this case here.}.

Unfortunately, for non-bi-diagonal solutions it is not as straightforward to generalize our results.
The main difficulty is that, unlike the case where the metrics are equal (or proportional), 
the matrix $\mathcal{K}^\mu_\nu$, expressed in terms of $h^\mu_\nu$, does not have a simple form like (\ref{K}).
Notice that for $f_{\mu\nu} = \omega^2g_{\mu\nu}$ each element of $\mathcal{K}^\mu_\nu$ depends only on 
$h^{\mu}_{\nu}$ (or $h^{(-)\mu}_{\phantom{(-)}\nu}$ in the theory with two dynamical metrics). 
This happens because in the case of proportional metrics, computing $K^\mu_\nu$ boils down 
to taking a square root of $(\delta^\mu_\nu + h^{\mu}_{\nu})$, 
resulting in $\delta^\mu_\nu + \frac12h^{\mu}_{\nu} + \mathcal{O}(h^2)$.
Instead for the non-bi-diagonal case the expression  $\sqrt{g^{\mu\alpha}_{(0)}f_{\alpha\nu}^{(0)} + h^{\mu}_{\nu}}$
must be evaluated, where $g^{\mu\alpha}_{(0)}f_{\alpha\nu}^{(0)}$ is {\it not} proportional to the unity matrix. An explicit analysis is needed to address the stability issue of such solutions.
We leave the non-bi-diagonal case for future investigations. 

We would like to stress that the instability rate $\Omega=\Omega(m')$ depends on the mass of the graviton.
As it was shown in {\cite{Emparan:2009at}}, the characteristic inverse time of the instability, for a fixed black hole mass $M=M_P^2r_S/2$, 
linearly grows as a function of the graviton mass for small $m'$, reaches a maximum 
at some $m'\sim r_S^{-1}$  and then decreases.
In most situations that are physically relevant $m\sim m' \ll r_S^{-1}$, since the graviton mass must be very small in order to satisfy solar system tests of gravity, say by the Vainshtein mechanism. In particular, to be responsible for the cosmic acceleration, one usually sets $m\sim H$, where $H$ is the Hubble parameter.
Therefore in the diagram of the instability \cite{Gregory:1993vy} typically we are sitting in the regime of small graviton masses, 
which means that the characteristic instability scale is of order of the inverse graviton mass, $\tau\sim m^{-1}$ and is independent of the black hole mass.
Therefore for $m\sim H$ the characteristic time of the instability is of the order of Hubble time:
for these scenarios, 
although the instability is present, it is not physically relevant for astrophysical black holes.\footnote{We point out that in principle for small $\kappa$ (but bigger than $(mr_S)^2$ to satisfy the inequality (\ref{inst})) the instability rate is enhanced by the factor $1/\sqrt{\kappa}$.} 

It is interesting to discuss the recent work~\cite{Mirbabayi:2013sva} in connection to our results, where
black holes in the dRGT theory of  massive gravity (with one dynamical metric, the other being fixed and equal to Minkowski) are considered. 
To get rid of the physical singularity of a bi-diagonal solution, the authors 
assumed time dependence of the physical metric, which results in the loss of 
mass by the black hole with the rate $\dot M \sim - r_S^2 m^2 M_P^2$. 
The ``instability'' scale associated with the process is $\tau \sim 1/(r_S m^2)$.
This time scale is much larger than the characteristic time of instability we found for $r_S \ll m^{-1}$.
This means that the Gregory-Laflamme type of instability is much more efficient than the one found in~\cite{Mirbabayi:2013sva}.
The difference in results must not be surprising, since both the physical processes leading to the instabilities 
and the black hole solutions studied in~\cite{Mirbabayi:2013sva} and here are very different.
The instability of black holes found in~\cite{Mirbabayi:2013sva} is due to the gradual flow of energy from the black hole, 
in fact it  can be seen as an accretion of ``phantom'' energy onto the black hole with associated decreasing 
of the black hole mass~\cite{Babichev:2004yx}.
The solution studied in~\cite{Mirbabayi:2013sva} contains the flat reference metric, whereas we consider background solutions 
with two metrics being identical to the Schwarzschild metric.
Moreover, the solution for the physical metric in~\cite{Mirbabayi:2013sva} features the Yukawa decay far from the black hole.  
As it was pointed out in~\cite{Mirbabayi:2013sva}, the instability of such black holes was already conjectured in~\cite{Dvali}.
On the contrary, the bi-Schwarzschild solutions we study contain no asymptotic exponential decay and can be considered as solutions 
in the Vainshtein regime all the way from the horizon to infinity.

There are several interesting questions that remain unanswered. 
First of all, the instability we discuss in this work operates only at linear level. 
It is not clear, however,  what would be the result of this instability after it grows to the nonlinear level, and, in general, 
what the fate of such unstable black holes is. 
It is worthwhile to mention that in the context of the black string instability in models with extra dimension(s), 
the linear instability grows to the nonlinear one, which results at the end in breaking up of the black string into an array of spherical black holes. 
The linear effective 4D description, however, ceases to work when the instability becomes nonlinear. 
Similarly, in bi-metric massive gravity we are not able to interpret the fate of bi-Schwarzschild black holes  without the nonlinear analysis. 
It could be a stable black hole solution, if it exists, or something else which is not a black hole. 
It was suggested in~\cite{Mirbabayi:2013sva}, for a completely different physical process, that the end of that instability is Minkowski spacetime. 
On the other hand the instability of the simplest (bi-Schwarzschild) black holes may indicate that  static black holes 
in massive gravity do not exist and/or that they do not form in the process of gravitational collapse. 

It would be also interesting to check if our results may be applied to other black hole solutions,
in particular to non-bi-diagonal solutions,  
and in other viable massive gravity theories, in particular in the Lorentz violating 
massive gravity theory with 5 propagating degrees of freedom, recently proposed in~\cite{Comelli:2013paa}.

To summarize, we studied perturbations on top of black hole solutions in dRGT massive gravity theory 
and its bi-gravity extension with two dynamical metrics. 
We found that the simplest black hole solutions, namely the bi-Schwarzschild solutions, are unstable.
Indeed,  the perturbation equations for such solutions coincide with those for the relevant perturbations of the 
higher-dimensional black string solutions featuring the well known Gregory-Laflamme instability. 
The instability time scale is of order of the graviton mass, therefore if the massive gravity is responsible for Dark Energy, 
then the instability is very mild, of the order of Hubble time. 

{\em Acknowledgments}.  We would like to thank Christos Charmousis, Marco Crisostomi and Renaud Parentani for useful discussions 
and Marco Caldarelli for correspondence. 
The work of E.B. was supported in part by grant FQXi-MGA-1209 from the Foundational Questions Institute.
A.F. thanks LPT Orsay for hospitality during various visits.

{\em Note added}: After our submission, another paper~\cite{Brito:2013wya} appeared in arXiv which confirms our conclusions on instability of black holes in massive gravity.

\end{document}